\documentclass[fleqn,twoside]{article}

\usepackage[headings]{espcrc2}

\readRCS
$Id: espcrc2.tex,v 1.2 2004/02/24 11:22:11 spepping Exp $
\usepackage{latexsym,bm,amsmath,amssymb,amsfonts}
    \usepackage{epsfig,graphics,graphicx}

\usepackage[figuresright]{rotating}

 \newcommand{\rmd}{{\rm d}}   

  \long\def\comment#1{ }
  \newcommand{\dif}{{\rm d}}
  \newcommand{\dY}{\dif Y}
  \newcommand{\abar}{\bar{\alpha}_s}
  
  \newcommand{\del}{\partial}

  \newcommand{\mcal}{\mathcal}
  \newcommand{\rme}{{\rm e}}

  \newcommand{\tr}{{\rm tr}}

  \newcommand{\Lam}{\Lambda_{{\rm QCD}}}
  
  \newcommand{\nn}{\nonumber\\}
   
  \newcommand{\order}[1]{\mcal{O}{(#1)}}

  \newcommand{\beq}{\begin{eqnarray}}
  \newcommand{\eeq}{\end{eqnarray}}
  
 \def\simge{\mathrel{%
   \rlap{\raise 0.511ex \hbox{$>$}}{\lower 0.511ex \hbox{$\sim$}}}}
\def\simle{\mathrel{
   \rlap{\raise 0.511ex \hbox{$<$}}{\lower 0.511ex \hbox{$\sim$}}}}

\title{Color Glass Condensate and its relation to HERA physics}

\author{Edmond Iancu\address{Institut de Physique Th\'eorique de Saclay,
 F-91191 Gif-sur-Yvette, France}}

\runtitle{Color Glass Condensate and HERA}
\runauthor{E. Iancu}

\begin{document}

\begin{abstract}
I give a brief overview of the effective theory for the Color Glass
Condensate, which is the high--density gluonic matter which controls
high--energy scattering in QCD in the vicinity of the unitarity limit. I
concentrate on fundamental phenomena, like gluon saturation,
unitarization, and geometric scaling, and the way how these are encoded
in the formalism. I emphasize the importance of the next--to--leading
order corrections, especially the running of the coupling, for both
conceptual and phenomenological issues. I survey the implications of the
CGC theory for the HERA physics and its phenomenological applications
based on saturation models.
 \vspace{1pc}
\end{abstract}

\maketitle

\section{Gluons at HERA}

\begin{figure*} [t]
\centerline{
 \includegraphics[width=0.45\textwidth]{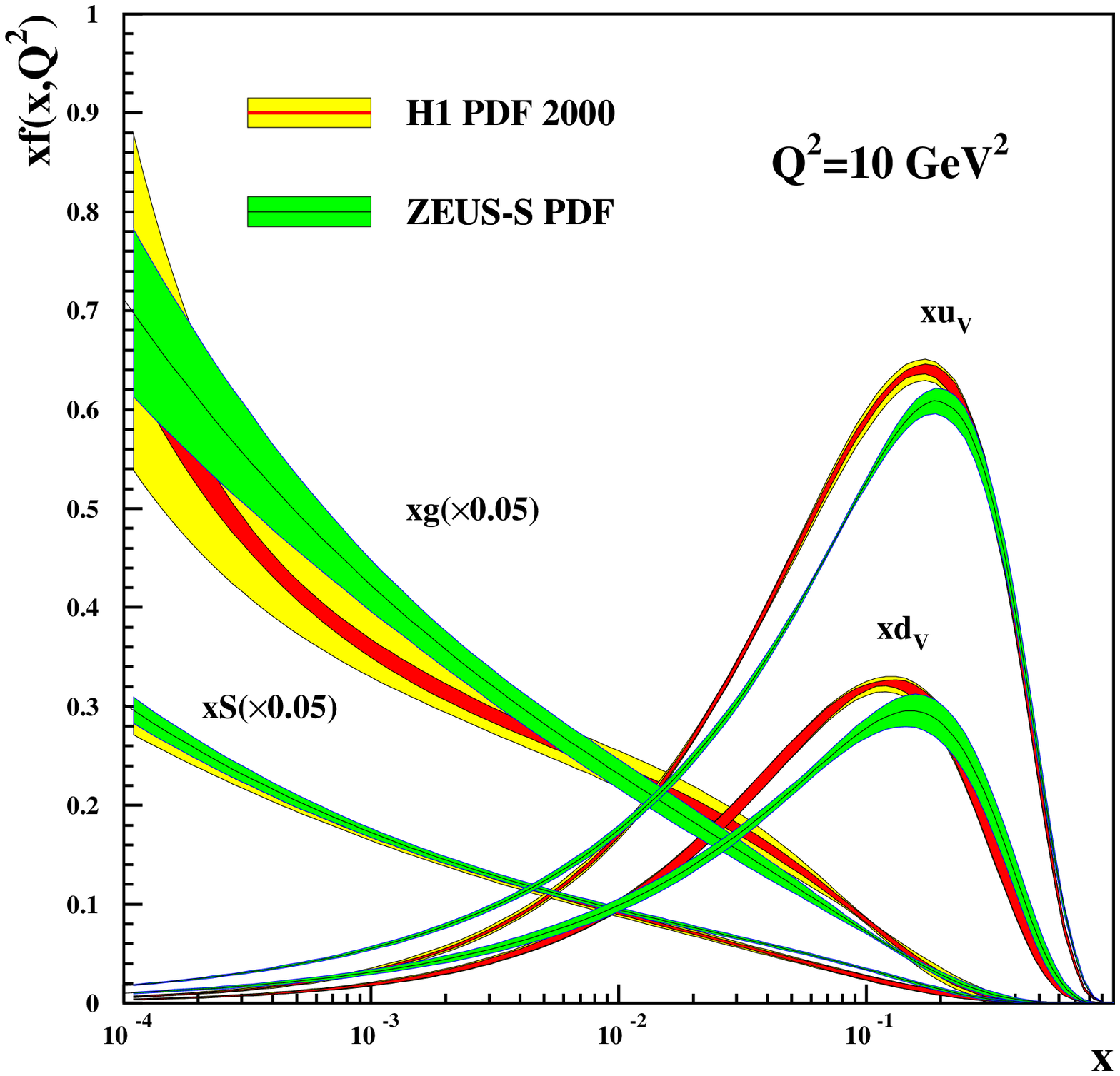}
 \includegraphics[width=0.5\textwidth]{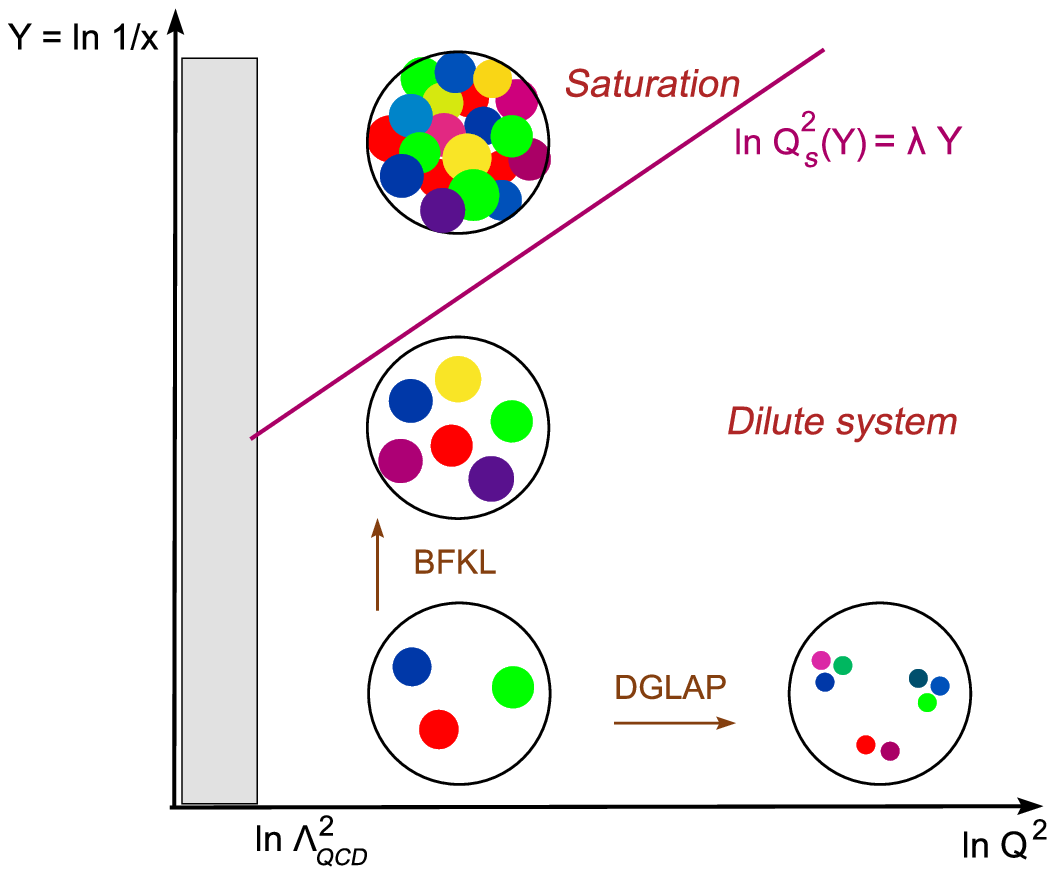}}\vspace*{-.6cm}
\caption{\sl Left: The $1/x$--evolution of the gluon, sea quark,
 and valence quark distributions for $Q^2=10$ GeV$^2$, as measured at HERA
 (combined fits to the H1 and ZEUS data). Note that the
 gluon and sea quark distributions have been reduced by a factor of 20 to
 fit inside the figure. Right: The
`phase--diagram' for QCD evolution as inferred from the HERA data; each
colored blob represents a parton with transverse area $\Delta x_\perp
\sim 1/Q^2$ and longitudinal momentum $k_z=xP_z$. \label{HERA-gluon}}
\end{figure*}

The essential observation at the basis of the recent theoretical progress
in the physics of hadronic interactions at high energy is the fact that
high--energy QCD is the realm of high parton (gluon) densities and hence
it can be studied from first principles, via weak coupling techniques.
Anticipated by theoretical developments like the BFKL equation
\cite{BFKL} and the GLR mechanism \cite{GLR,MQ85} for gluon saturation,
this observation has found its first major experimental foundation in the
HERA data for electron--proton deep inelastic scattering (DIS) at
small--$x$. As visible in the leftmost figure in Fig. \ref{HERA-gluon},
the gluon distribution $xG(x,Q^2)$ rises very fast when decreasing
Bjorken--$x$ at fixed $Q^2$ --- roughly, as a power $1/x^\lambda$ with
$\lambda\simeq 0.2\div 0.3$. The physical interpretation of such results
is most transparent in the proton infinite momentum frame, where
$xG(x,Q^2)$ is simply the number of the gluons in the proton wavefunction
which are localized within an area $\Delta x_\perp \sim 1/Q^2$ in the
transverse plane and carry a fraction $x=k_z/P_z$ of the proton
longitudinal momentum.

Thus, without any theoretical prejudice, the HERA data suggest the
physical picture illustrated in the right hand side of Fig.
\ref{HERA-gluon}, which shows the distribution of partons in the
transverse plane as a function of the kinematical variables for DIS in
logarithmic units: $\ln Q^2$ and $Y\equiv \ln(1/x)$. The number of
partons increases both with increasing $Q^2$ and with decreasing $x$, but
whereas in the first case (increasing $Q^2$) the transverse area $ \sim
1/Q^2$ occupied by every parton decreases very fast and more than
compensates for the increase in their number --- so, the proton is driven
towards a regime which is more and more dilute ---, in the second case
(decreasing $x$) the partons produced by the evolution have roughly the
same transverse area, hence their density is necessarily increasing.

Accordingly, the DGLAP equation \cite{DGLAP} which describes the
evolution with increasing $Q^2$ is naturally {\em linear}, and also {\em
local} in $Q^2$. By contrast, the BFKL equation, which is the linear
equation originally proposed \cite{BFKL} to describe the evolution with
increasing energy, is {\em non--local} in transverse space and should be
merely regarded as a linear approximation to more general evolution
equations which are {\em non--linear}, i.e., which account for the
interactions among the partons within the wavefunction. The non--linear
effects are expected to become important in the region denoted as
`saturation' in Fig. \ref{HERA-gluon}, and in the approach towards it
from the dilute region at large $Q^2$.

Mainly because of its complexity, the high--energy evolution in QCD is
not as precisely known as the corresponding evolution with $Q^2$. Still,
the intense theoretical efforts over the last years led to important
conceptual clarifications and to new, more powerful, formalisms --- among
which, the effective theory for the Color Glass Condensate
(CGC)~\cite{MV,JKLW,CGC,B,K,EdiCGC}
---, which encompass the non--linear dynamics in high--energy QCD
to lowest order in $\alpha_s$ and allow for a unified picture of
various high--energy phenomena ranging from DIS to heavy--ion, or
proton--proton, collisions, and to cosmic rays.

These developments provide a natural explanation for a variety of
remarkable phenomena observed in the current experiments, like the
`geometric scaling' in the HERA data at small $x$ \cite{geometric,MS06}
and the particle production at forward rapidities in deuteron--gold
collisions at RHIC \cite{Brahms-data}. Moreover, they have potentially
interesting predictions for the physics at LHC. It is our purpose in what
follows to provide a brief, pedagogical, introduction to such new ideas,
with emphasis on the physical picture and its consequences for deep
inelastic scattering at high energy.

\section{DIS in the dipole frame}
\label{sec:dipole}

At small $x$, DIS is most conveniently computed by using the {\em dipole
factorization} (see, e.g., Refs.~\cite{EdiCGC} for more details and
references). The small--$x$ quark to which couple the virtual photon is
typically a `sea' quark produced at the very end of a gluon cascade. It
is then convenient to disentangle the electromagnetic process
$\gamma^*q$, which involves this `last' emitted quark, from the QCD
evolution in the proton, which involves mostly gluons. This can be done
via a Lorentz boost to the `dipole frame' in which the struck quark
appears as an excitation of the virtual photon, rather than of the
proton. In this frame, the proton still carries most of the total energy,
while the virtual photon has just enough energy to dissociate long before
the scattering into a `color dipole' (a $q\bar q$ pair in a color singlet
state), which then scatters off the gluon fields in the proton. This
leads to the following factorization:
  \beq\label{dipolefact}
  \sigma_{\gamma^*p}(x,Q^2) \!\!
 & = & \!\!\int_0^1 \rmd z \int \rmd^2 r\ \vert \Psi_{\gamma}(z,r;
 Q^2)\vert^2 \qquad\nn &{}&\qquad \quad \times \ \sigma_{\rm dipole}(x,r)
 \eeq
where $\vert \Psi_{\gamma}(z,r; Q^2)\vert^2$ is the probability for
the $\gamma^*\to q\bar q$ dissociation ($r$ is the dipole transverse
size and $z$ the longitudinal fraction of the quark), and
$\sigma_{\rm dipole}(x,r)$ is the total cross--section for
dipole--proton scattering and represents the hadronic part of DIS.
At high energy, the latter can be computed in the eikonal
approximation as
 \beq\label{sigdip}  \sigma_{\rm dipole}(x,r)\ = \ 2\int \rmd^2b\ \ T(x,r,b)
 \,\eeq
where $T(x,r,b)$ is the {\em forward scattering amplitude} for a dipole
with size $r$ and impact parameter $b$. This is the quantity that we
shall focus on. The unitarity of the $S$--matrix requires $T\le 1$, with
the upper limit $T=1$ corresponding to total absorbtion, or `black disk
limit'.

But the unitarity constraint can be easily violated by an incomplete
calculation, as we demonstrate now on the example of lowest--order (LO)
perturbation theory. To that order, $T(r,b,Y)$ involves the exchange of
two gluons between the dipole and the target. Each exchanged gluon brings
a contribution $gt^a \bm{r}\cdot\bm{E}_a$, where $\bm{E}_a$ is the color
electric field in the target. Thus, $T\sim g^2 r^2 \langle
\bm{E}_a\cdot\bm{E}_a\rangle_x$, where the expectation value in the
r.h.s. is recognized as the number of gluons per unit transverse area, as
measured by a probe with transverse resolution $Q^2\sim 1/r^2$ :
  \beq\label{T1SCATT}
    T(x,r,b) \,\sim\,
  \alpha_s\,r^2\,\frac{xG(x,1/r^2)}{\pi
  R^2}\,\equiv\, \alpha_s\,n(x, Q^2)\,.\eeq
Above, we identified the {\em gluon occupation number}\,: $n(x,Q^2)=$
[number of gluons $xG(x,Q^2)$] times [the area $1/Q^2$ occupied by each
gluon] divided by [the proton transverse area $\pi R^2$].

Eq.~(\ref{T1SCATT}) applies so long as $T\ll 1$ and shows that weak
scattering (or `color transparency') corresponds to low gluon
occupancy $n\ll 1/\alpha_s$. But if naively extrapolated to very
small values of $x$, this formula leads to {\em unitarity
violations} : $T$ would eventually become larger than one ! Before
this happens, however, new physical phenomena are expected to come
into play and restore unitarity. As we shall see, these are {\em
non--linear} phenomena, and are of two types: \texttt{(i)} {\em
multiple scattering}, i.e., the exchange of more than two gluons
between the dipole and the target, and \texttt{(ii)} {\em gluon
saturation}, i.e., non--linear effects in the proton wavefunction
which tame the rise of the gluon distribution at small~$x$.

Eq.~(\ref{T1SCATT}) also provides a criterion for the onset of unitarity
corrections: these become important when $T(x,r)\sim 1$ or $n(x,Q^2)\sim
1/\alpha_s$. This condition can be understood as follows: to have
non--linear phenomena, the gluons in the wavefunction must be numerous
enough (which requires small $x$) and large enough (meaning low
transverse momenta $Q^2$) in order to strongly overlap with each other,
by a factor $n\sim {1/\alpha_s}\gg 1$ which is large enough to compensate
for the smallness of the coupling. When this happens, the gluon mutual
interactions $\sim n\alpha_s$ become of $\order{1}$.

The condition $n(x,Q^2)\sim 1/\alpha_s$ can be solved for the {\em
saturation momentum}, which is the value of the transverse momentum below
which saturation effects are expected to be important in the gluon
distribution. One thus finds
 \begin{equation}\label{Qsat}
   Q^2_s(x)\, \simeq \,
{\alpha_s }\, \frac{x G(x,Q^2_s)}{\,\pi R^2}\, \,\,\sim\,\,
 \frac{1}{x^{\lambda}}\,,\end{equation}
which grows with the energy as a power of $1/x$, since so does the gluon
distribution before reaching saturation. In logarithmic units, with
$Y\equiv \ln(1/x)$, the {\em saturation line} $\ln Q^2_s(Y)=\lambda Y$ is
therefore a {\em straight} line, as illustrated in the right hand side of
Fig. \ref{HERA-gluon}. This is the borderline between the dilute regime
at high transverse momenta $k_\perp\gg Q_s(Y)$, where one expects the
standard perturbation theory to apply, and a high--density region at low
momenta $k_\perp\simle Q_s(Y)$, where physics is non--linear. In fact, as
we shall argue below, at high energy the effects of saturation can extend
up to very high values of $k_\perp$, well above the saturation line.

\section{BFKL evolution}
\label{sec:BFKL}

\begin{figure}
\begin{center}
\centerline{\epsfig{file=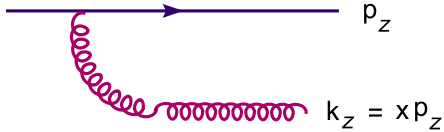,height=1.8cm}}\vspace*{-.6cm}
 \caption{\sl
Bremsstrahlung at lowest order.\label{OneGluon}}
\end{center}\vspace*{-1.cm}
\end{figure}

Within perturbative QCD, the emission of small--$x$ gluons is amplified
by the infrared sensitivity of the bremsstrahlung process, whose
iteration leads to the BFKL evolution (at least, for not too high
energies). Fig.~\ref{OneGluon} shows the emission of a gluon which
carries a fraction $x=k_z/p_z$ of the longitudinal momentum of its parent
quark. When $x\ll 1$, the differential probability for this emission can
be estimated as
 \beq\label{brem} \rmd P_{\rm Brem}\,\simeq\,\frac{\alpha_s
 C_F}{2\pi^2}\,\frac{\rmd^2k_\perp}{k_\perp^2}\,\frac{\rmd
 x}{x}\,,\eeq
which is singular as $x\to 0$.
Introducing the rapidity $Y=\ln(1/x)$, and hence $\rmd Y=\rmd x/x$,
Eq.~(\ref{brem}) shows that there is a probability of $\order{\alpha_s}$
to emit one gluon per unit rapidity. The same would hold for the emission
of a soft photon from an electron in QED. However, unlike the photon, the
child gluon is itself charged with `colour', so it can further emit an
even softer gluon, with longitudinal fraction $x_1=q_z/k_z\ll 1$. When
the rapidity is large, $\alpha_s Y\gg 1$, such successive emissions lead
to the formation of gluon cascades, in which the gluons are ordered in
rapidity and which dominate the small--$x$ part of the hadron
wavefunction.

\comment{\begin{figure*}[t]
\begin{center}
\centerline{\epsfig{file=Evol4.eps,width=0.8\textwidth}}
\caption{\sl Gluon cascades produced by the high--energy (BFKL)
evolution of the proton wavefunction, as probed by a small color
dipole in DIS.\label{BFKLfig}}
\end{center}\vspace*{-.8cm}
\end{figure*}}

So long as the density is not too high, the gluons do not interact
with each other and the evolution remains {\em linear} : when
further increasing the rapidity in one more step ($Y\to Y+\dY$), the
gluons created in the previous steps {\em incoherently}  act as {\em
color sources} for the emission of a new gluon. This picture leads
to the following, schematic, evolution equation
 \beq\label{BFKL0}
 \frac{\partial n}{\partial Y}\,\simeq\,\omega\alpha_s n \quad
 \Longrightarrow \quad n(Y)\,\propto\, {\rm e}^{\omega\alpha_s Y}\,,
 \eeq
which predicts the exponential rise of $n$ with $Y$. There is an
additional feature of the high energy evolution which needs to be
emphasized: the gluon emission vertex is non--local in transverse
momentum (the transverse momentum of the daughter gluon is generally
different from that of its parent), but this non--locality is quite weak
and can be described as {\em diffusion} in the logarithmic momentum
variable $\rho\equiv\ln Q^2$. That is, a better version of
Eq.~(\ref{BFKL0}) reads
 \beq\label{BFKL}
 \frac{\partial n(Y,\rho)}{\partial Y}\,\simeq\,\omega\alpha_s n
 \,+\,\chi\alpha_s \partial_\rho^2 n\,,
 \eeq
where $\chi$ is the diffusion coefficient. This is an oversimplified
version of the BFKL (Balitsky-Fadin-Kuraev-Lipatov) equation \cite{BFKL}
which captures the main features of this evolution: the unstable growth
of the gluon distribution and the diffusion in transverse momentum. Both
features leads to difficulties in the high--energy limit: \texttt{(i)}
the unlimited growth of $n$ entails unitarity violations, as discussed in
the previous section; \texttt{(ii)} the BFKL evolution explores the
transverse phase--space in a diffusive way, so its solution $n(Y,\rho)$
receives significant contributions from all the points $\rho'$ such that
$|\rho'-\rho|\lesssim\sqrt{\chi\alpha_sY}$. Hence, even if $\rho$ is
hard, the momenta $\rho'$ contributing to the solution can be
considerably softer; with increasing $Y$, there is a larger and larger
part of the total result which is generated from soft momenta (on the
left of the saturation line), where linear evolution, or even
perturbation theory, are bound to fail. One knows by now that both
features --- the exponential growth of $n$ with $Y$ and the BFKL
diffusion --- are considerably tempered by NLO effects
\cite{NLBFKL,Salam99}, like the running of the coupling or the constraint
of energy conservation. But the basic fact that the gluon density
increases very fast with $Y$ is expected to remain true (to all orders in
$\alpha_s$) so long as one neglects the {\em non--linear} effects in the
evolution.

\section{JIMWLK evolution and the
CGC}

\begin{figure*}
 \begin{center}
 \centerline{\epsfig{file=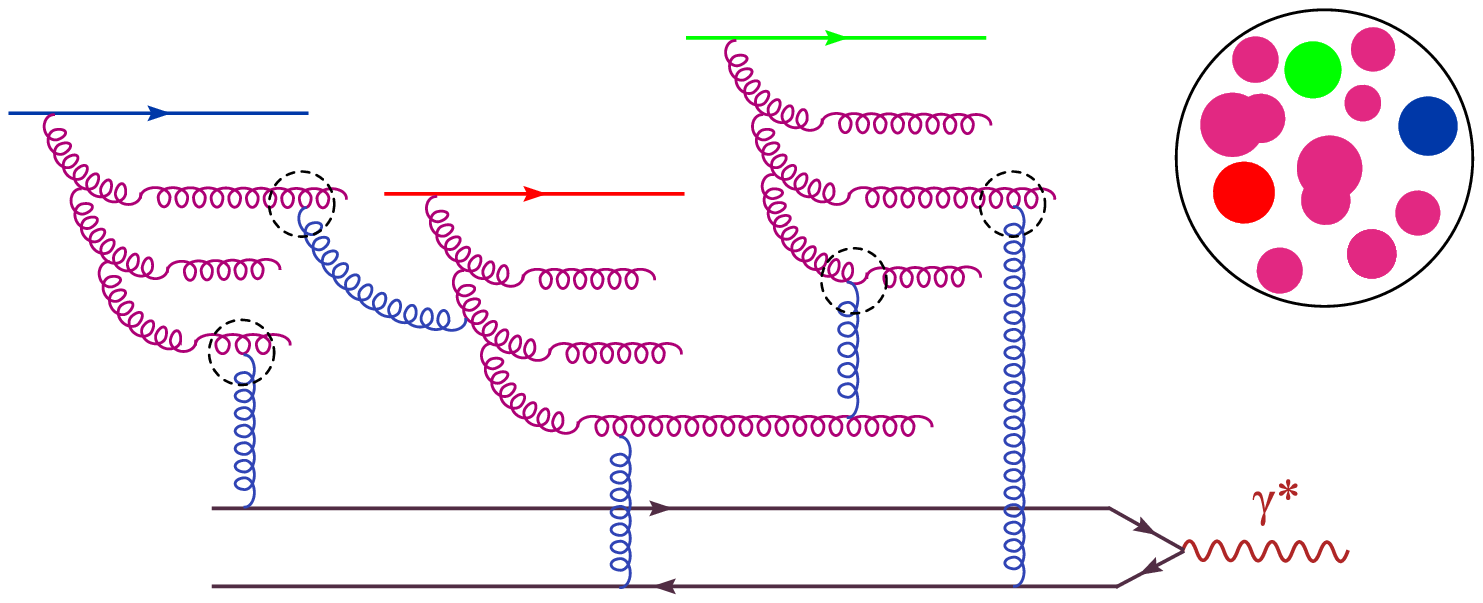,width=0.8\textwidth}}
   \caption{\sl  DIS in the presence of BFKL evolution, saturation
   and multiple scattering.\label{EvolREC}}
            \end{center}\vspace*{-1.cm}
\end{figure*}

Non--linear effects appear because gluons carry colour charge,
so they can interact with each other 
(even when separated in rapidity) by exchanging gluons in the
$t$--channel, as illustrated in Fig. \ref{EvolREC}. These
interactions are amplified by the gluon density and thus they should
become more and more important when increasing the energy.

Back in 1983, L. Gribov, Levin and Ryskin \cite{GLR} (see also Ref.
\cite{MQ85}) suggested that gluon saturation should proceed via $2\to 1$
`gluon recombination', which is a process of order $\alpha_s^2 n^2$ (cf.
Fig. \ref{EvolREC}). One can heuristically take this into account by
adding a non--linear term to the r.h.s. of Eq.~(\ref{BFKL}) :
 \beq\label{GLR}
  \frac{\partial n}{\partial Y}\,\simeq\,\alpha_s \partial_\rho^2 n
  +\alpha_s n-
{\alpha_s^2} n^2.
 \eeq
Clearly, this non--linear equation has a {\em fixed point} $n_{\rm
sat}\sim 1/\alpha_s$ at high energy. That is, when $n$ is as high as
$1/\alpha_s$, the emission processes (responsible for the BFKL growth)
are precisely compensated by the recombination ones, and then the gluon
occupation factor saturates at a fixed value.

Twenty years later, we know that the actual mechanism for gluon
saturation in QCD is more subtle than just gluon recombination and that
its mathematical description is considerably more involved than suggested
by Eq.~(\ref{GLR}). This mechanism, as encoded in the effective theory
for the CGC and its central equation, the JIMWLK equation
(Jalilian-Marian, Iancu, McLerran, Weigert, Leonidov, and Kovner)
\cite{JKLW,CGC}, is the {\em saturation of the gluon emission rate due to
high density effects} : At high density, the gluons are not independent
color sources, rather they are correlated with each other in such a way
to ensure {\em color neutrality }\cite{SAT,AM02,GAUSS} over a distance
$\Delta x_\perp \sim 1/Q_s$. Accordingly, the soft gluons with
$k_\perp\simle Q_s$ are {\em coherently} emitted from a quasi--neutral
gluon distribution, and then the emission rate ${\partial n}/{\partial
Y}$ saturates at a constant value of $\order{1}$. Thus, in the regime
that we call `saturation', the gluon occupation factor keeps growing, but
only {\em linearly} in $Y$ (i.e., as a logarithm of the energy)
\cite{AM99,SAT}. Schematically:

 \beq\label{dndy}
 \frac{\partial n}{\partial Y}\,=\,\chi(n)\,\approx
 \begin{cases}
        \displaystyle{\alpha_s n\,} &
        \text{ if\,  $n\ll {1}/{\alpha_s}$ }
        \\*[0.35cm]
        \displaystyle{\ 1} &  
        \text{ if\,  $n\simge {1}/\alpha_s$}
    \end{cases}
  \eeq
where $\chi(n)$ is a non--linear function with the limiting behaviours
displayed above. The transition between the two regimes is smooth and it
occurs around the saturation line, i.e., at transverse momenta
$k_\perp\sim Q_s(Y)$, where $Q_s(Y)$ is an increasing function of $Y$
which is determined by the theory. As we shall later explain, this rise
is roughly consistent with the power--like increase with $1/x$ predicted
by Eq.~(\ref{Qsat}).

Eq.~(\ref{dndy}) is not yet the JIMWLK equation, but only a mean field
approximation to it: in reality, one cannot write down a closed equation
for the 2--point function $n(Y)=\langle \bm{E}_a\cdot\bm{E}_a\rangle_Y$,
but only an {\em infinite hierarchy} for the $N$--point correlations
$\langle A(1)A(2)\cdots A(N)\rangle_Y$ of the gluon fields. This is so
since the $N$--point functions couple under the evolution via the
non--linear effects. In the CGC formalism, these correlations are encoded
into the {\em weight function} $W_Y[A]$
--- a functional probability density for the color field configurations:
 \beq\label{W}
  \langle A(1)\, A(2)\, \cdots\,A(N) \rangle_Y\,=\qquad\qquad
  \qquad\qquad\nn
    =\int {\rm D}[A]\ W_Y[A]\ A(1)\, A(2)\, \cdots\, A(N)\,.
  \eeq
The average in Eq.~(\ref{W}) is similar to the `average over disorder'
that is usually performed in the study of amorphous materials, like
glasses: the various target configurations scatter independently with the
incoming projectile (indeed, their internal dynamics is `frozen' over the
characteristic time scale for scattering, by Lorentz time dilation), and
the physical scattering amplitude is finally obtained by summing the
contributions from all such configurations, with weight function
$W_Y[A]$. This explains the concept of `glass' in the `Color Glass
Condensate'. The `color' refers, of course, to the gluon color charge.
Finally, the `condensate' stays for the coherent state made by the gluons
at saturation: this state has a large occupation number $n\sim
1/\alpha_s\gg 1$, as typical for a Bose condensate.

The JIMWLK equation \cite{JKLW,CGC} is a {\em functional} differential
equation describing the evolution of $W_Y[A]$ with $Y$. Via
Eq.~(\ref{W}), this functional equation generates an infinite hierarchy
of ordinary evolution equations for the $N$--point functions of $A$, as
anticipated. To describe physical observables, these correlations must be
gauge--invariant. At high energy, a particularly convenient set of
gauge--invariant correlation functions is obtained by taking products of
{\em Wilson lines} traced over the color indices.

A `Wilson line' describes the scattering between a high energy parton and
a gauge background field in the eikonal approximation: the parton
preserves a straight--line trajectory while moving through the field.
Hence, the product of $N$ Wilson lines describes the $S$--matrix for $N$
partons propagating in the background field $A$. After also averaging
over $A$, as in Eq.~(\ref{W}), one finally obtains the $S$--matrix for
the eikonal scattering between the partonic system and the hadron (the
`CGC'). For instance, the average $S$--matrix for dipole scattering, as
relevant to DIS (cf. Fig. \ref{EvolREC}), is computed as
 \beq\label{Sdipole} \langle
S({\bm r}, {\bm b})\rangle_Y=\int {\rm D}[A]\ W_Y[A]\ \frac{1}{N_c}\, \tr
 (V^\dagger_{\bm x} V_{\bm y})\eeq
where $V^\dagger_{\bm x}$ is the Wilson line for the quark with
transverse coordinate ${\bm x}$, i.e.,
 \beq\label{Wilson}
  V^\dag(\bm x)={\mbox P}\exp \left( ig\int dx^-
  A^+_a(x^-,\bm x)t^a \right)\,,
  \eeq
$V_{\bm y}$ is the corresponding operator for the antiquark at ${\bm y}$,
and the dipole size and impact parameter are given by ${\bm r = \bm x-
\bm y}$, ${\bm b = (\bm x+ \bm y)/2}$. The (average) forward scattering
amplitude, which is the quantity which determines the dipole
cross--section (\ref{sigdip}), is then obtained as $\langle T\rangle = 1-
\langle S\rangle$.

In this description, the {\em unitarity corrections} are explicit: the
multiple scattering is encoded in the Wilson lines and the gluon
saturation in the weight functional $W_Y[A]$. Hence, no surprisingly, the
corresponding evolution equations, as deduced from the JIMWLK equation,
have the property to manifestly preserve unitarity. These equations form
an infinite hierarchy which was originally derived (within a different
formalism) by Balitsky \cite{B}. The first equation in this hierarchy
reads (with transverse coordinates omitted)
  \beq\label{B1}
  \del_Y\langle T\rangle\,=\,\alpha_s\langle T\rangle \,-\,
 \alpha_s\langle T^2\rangle\,.\eeq
This is not a closed equation: the amplitude $\langle T\rangle$ for the
scattering of one dipole is related to the amplitude $\langle T^2\rangle$
for two dipoles. A closed equation, known as the Balitsky--Kovchegov (BK)
equation \cite{K}, can be obtained in a mean field approximation which
assumes factorization: $\langle T^2\rangle\approx \langle T\rangle
\langle T\rangle$. After restoring the transverse coordinates in the
diffusion approximation, the BK reads (with $\rho\equiv \ln(1/r^2)$ and
$T\equiv \langle T\rangle$)
 \beq\label{BK}
  \del_Y T(Y,\rho)\,=\,
  \alpha_s \partial_\rho^2 T
  +\alpha_s T-{\alpha_s} T^2.
 \eeq
As anticipated, unitarity ($T\le 1$) is manifest on this equation which
has $T=1$ as a fixed point at high energy. Formally, the BK equation
represents the large--$N_c$ limit of the Balitsky--JIMWLK hierarchy.
Recent numerical studies demonstrate that this mean field aproximation
works better than expected: for $N_c=3$, the differences between the BK
and JIMWLK predictions for the dipole amplitude are less than 1\%
\cite{Kovchegov:2008mk}. These properties, together with the relative
simplicity of Eq.~(\ref{BK}), make this equation a very convenient tool
for studies of saturation and unitarity. Some of its physical
consequences will be described in the next sections.

Let us conclude this section on the general formalism with a few
additional remarks:

\texttt{(i)} The original motivation for the CGC theory \cite{MV} came
from the physics of ultrarelativistic heavy ion collisions, where this
theory is meant to describe the initial states of the incoming nuclei,
prior to collision. In that context, the gluon density is large already
at moderate energies, due to the existence of many `tree--level' color
sources: the $3A$ valence quarks, with $A$ the atomic number. The CGC
theory produced some interesting predictions for particle production in
nucleus--nucleus collisions, which have been since confirmed by the
experimental results at RHIC (see the review articles
\cite{EdiCGC,JalilianMarian:2005jf,Gelis:2007kn,CGCRHIC} and Refs.
therein). With the advent of LHC, the CGC theory should find a vaste
field of applications to both nucleus--nucleus and proton--proton
collisions \cite{Abreu:2007kv}.

\texttt{(ii)} The BK equation in the diffusion approximation, as written
down in Eq.~(\ref{BK}), turns out to be the same as the FKPP equation
which describes the mean field limit (corresponding to very large
occupation numbers) of a classical stochastic process known as {\em
reaction--diffusion}. This process can be briefly described as follows
\cite{Saar}: `molecules' of type $A$ which are located at the sites of an
infinite, one--dimensional, lattice can locally split ($A\to AA$) or
merge  with each other ($AA\to A$); also, a molecule can diffuse to the
adjacent sites. The correspondence between BK and FKPP, originally
noticed in Ref. \cite{MP03}, sheds new light on the physics of geometric
scaling and helps clarifying the limitations of the mean field
approximation (see below).

\texttt{(iii)} The central equations of the CGC formalism, so like the
Balitsky--JIMWLK hierarchy, or the factorization formula (\ref{Sdipole})
for DIS together with the corresponding ones for proton--nucleus
\cite{KM98,CM04}, or nucleus--nucleus \cite{MV,Gelis:2008rw}, collisions,
are known so far only to {\em leading order} (LO) in perturbation theory
--- an approximation which by itself involves an infinite resummation of
perturbative contributions where the powers of $g$ are accompanied by
appropriate powers of $Y=\ln s$, or of the strong gauge fields $A\sim
1/g$, so like in Eq.~(\ref{Wilson}). However, the higher--order effects,
and especially the running of the coupling, turn out to be extremely
important --- whenever known, they dramatically affect the predictions of
the LO theory, as we shall shortly see. Fortunately, there is an ongoing
effort towards the inclusion of higher--order corrections, which so far
has led to an improved version of the BK equation containing running
coupling effects \cite{Balitsky:2006wa,Kovchegov:2006vj}.

\texttt{(iv)} Even at leading order, the equations previously mentioned
(Balitsky--JIMWLK and BK) are still incomplete: they neglect the effects
of {\em gluon number fluctuations} (or `Pomeron loops'), i.e., the
correlations associated with the fact that some of the gluons produced by
the high--energy evolution have a common ancestor \cite{MS04,IMM04,IT04}.
However, in practice, this is not a serious drawback, since these
correlations are anyway suppressed by the running of the coupling
\cite{Dumitru:2007ew}. This will be further discussed in the next
section.

\section{Saturation and Geometric scaling}
\label{subsec:GEOM}

\begin{figure}[h]
    \begin{center}
             \epsfig{file=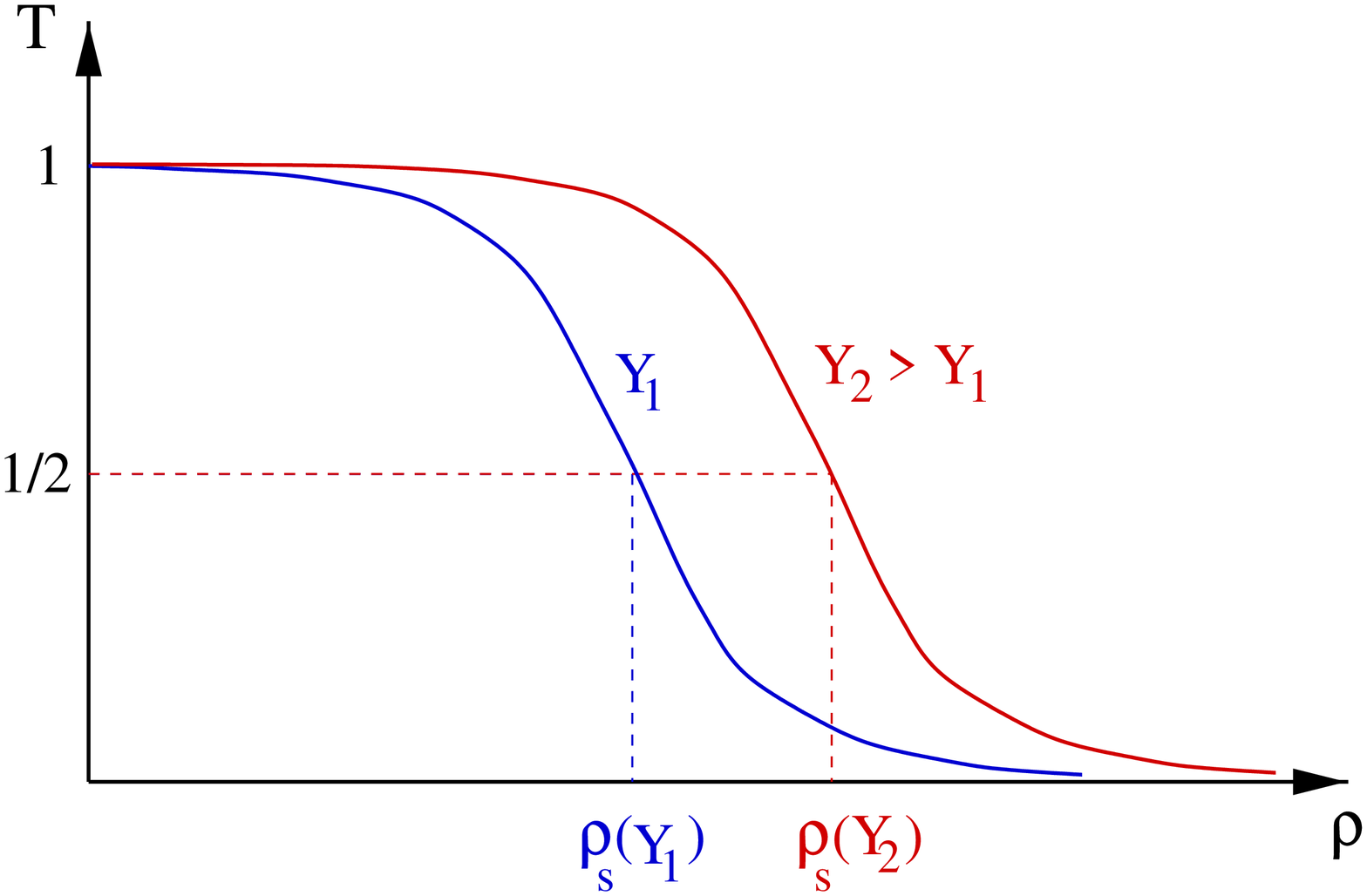,
        width=0.45\textwidth}\vspace*{-.6cm}
            \caption{\sl The saturation front generated by the BK equation
            for two values of the rapidity.}\end{center}
            \label{Front}
            \vspace*{-1.cm}
            \end{figure}

Due to its simplicity, the BK equation (\ref{BK}) is well suited for both
numerical and analytic studies of the evolution towards unitarity and
saturation. The corresponding solution $T(Y,\rho)$ is a {\em front} which
with increasing $Y$ propagates towards larger vales of $\rho$, in such a
way that the position of the front coincides with the saturation momentum
$\rho_s(Y)\equiv \ln Q_s^2(Y)$ \cite{SCALING,MT02,MP03} (see
Fig.~\ref{Front}). Behind the front, i.e., for $\rho< \rho_s(Y)$, the
dipole scattering amplitude has reached the black disk limit $T=1$
(which, we recall, is a fixed point of the BK equation) and hence it
cannot grow anymore. Ahead of the front ($\rho\gg \rho_s(Y)$), the
amplitude is still weak, $T\ll 1$, so the non--linear term in
Eq.~(\ref{BK}) is unimportant and the amplitude can grow according to the
linear, BFKL, evolution.

This argument suggests that the progression of the front with increasing
$Y$ is entirely driven by the linearized version of the BK equation
 --- a conclusion which is confirmed by more rigorous
mathematical arguments \cite{MP03}. Hence, by solving the linear, BFKL,
equation supplemented with a saturation boundary condition (namely,
$T(Y,\rho)\simeq 1$ for $\rho=\rho_s(Y)$), one can compute the position
$\rho_s(Y)$ of the front, which is the same as the {\em saturation line},
and also the shape of the amplitude ahead of the front. One thus finds
\cite{SCALING,MT02} that the front progresses at constant speed, i.e.,
the saturation line is a straight line (below, $\abar\equiv \alpha_s
N_c/\pi$) :
 \beq\label{rhos} \rho_s(Y)\,\simeq\,\lambda_s Y\quad{\rm with}\quad
 \lambda_s\approx 4.88\abar\,.\eeq
Moreover, within a relatively large window at $\rho> \rho_s(Y)$, whose
width is growing with $Y$, the amplitude depends upon $\rho$ and $Y$ only
via the difference $\tau\equiv \rho-\rho_s(Y)=\ln(1/r^2Q_s^2(Y))$ :
 \beq \label{scaling} T(Y,\rho)\propto \tau \,\rme^{
-\gamma_s\tau} \quad{\rm for}\quad 1<\tau
 < \sqrt{\chi\abar Y}\,,\eeq
with $\gamma_s\approx 0.63$. This property is known as {\em geometric
scaling}. This scaling also holds (trivially !) behind the front, since
$T$ is constant there: $T=1$ for $\rho\lesssim \rho_s(Y)$. Geometrically,
this means that, with increasing $Y$, the saturation front gets simply
translated towards larger values of $\rho$, but its shape remains
unchanged: the front propagates like a {\em traveling wave} \cite{MP03}.

We conclude that the dipole amplitude shows geometric scaling for all
values of $\rho$ up to a maximal value $\rho_{\rm geom}(Y)\simeq
\rho_s(Y)+\sqrt{\chi\abar Y}$, which for large $Y$ can be significantly
larger than $\rho_s(Y)$. (The difference $\rho_{\rm geom}- \rho_s\propto
\sqrt{Y}$ grows with $Y$ via BFKL diffusion.) This means that the effects
of saturation make themselves felt even at relatively large momenta $Q^2
\gg Q_s^2(Y)$, where the scattering is weak, $T\ll 1$, and the gluon
density in the target is quite low. This considerably extends the
phase--space where saturation is expected to be important in the
experiments.

\begin{figure}
\begin{center}\hspace*{-.1cm}
       \epsfig{figure=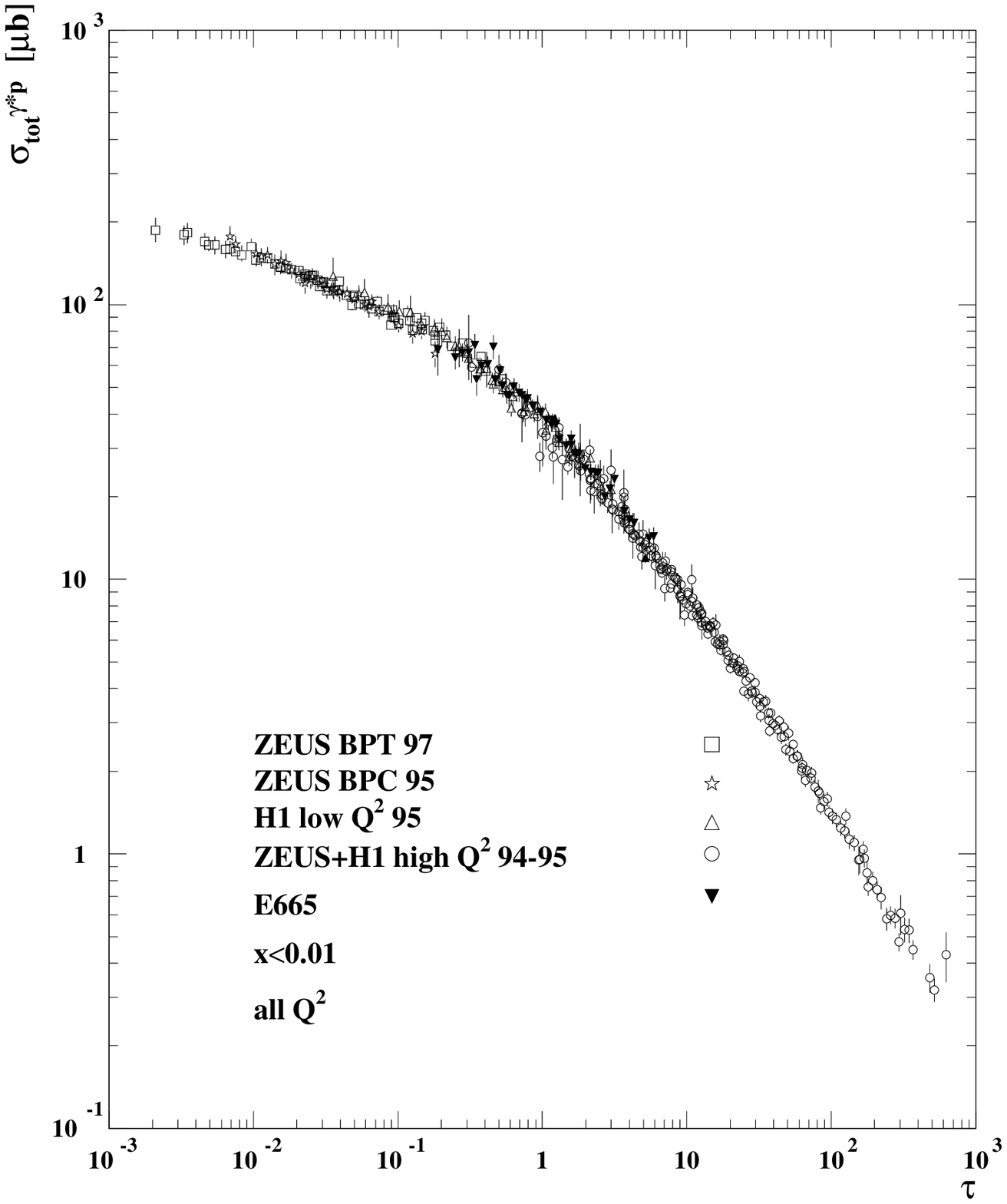,
        width=0.48\textwidth}    \vspace*{-.8cm}
\caption{\sl Geometric scaling in the HERA data for $\sigma_{\gamma^*p}$
at $x\le 0.01$; $\tau$ is the scaling variable, $\tau\equiv
Q^2/Q^2_s(Y)$ \cite{geometric}.} \label{HERA-scaling}
\end{center}\vspace{-1.cm}
\end{figure}

Via the factorization formula (\ref{dipolefact}), the scaling of the
amplitude as a function of $rQ_s(Y)$ implies a similar scaling for the
DIS cross--section: $\sigma_{\gamma^*p}(Y,Q^2)\approx\sigma({\tau})$,
with  $\tau\equiv Q^2/Q^2_s(Y)$. Such a scaling has been indeed
identified in the HERA data, by Sta\'sto, Golec-Biernat and Kwieci\'nski
\cite{geometric} (see Fig. \ref{HERA-scaling}). More recently, with the
advent of more precise data for DIS diffraction at HERA, geometric
scaling has been noticed in these data too \cite{MS06}. The data also
show {\em violations} of geometric scaling, which can be understood as
consequences of the BFKL diffusion \cite{IIM03} and of the quark masses
\cite{GS07}. These phenomenological studies will be further described in
the next section.

But the results at HERA also show that the rise of the gluon distribution
with $1/x$ is much slower than predicted by the LO BFKL analysis: when
described in terms of saturation, they require a saturation exponent
$\lambda_s = 0.2\div 0.3$ whereas Eq.~(\ref{rhos}) yields $\lambda_s
\simeq 1$ for $\alpha_s\simeq 0.2$. This discrepancy is solved by the NLO
calculation of the saturation exponent \cite{DT02} which predicts indeed
$\lambda_s\simeq 0.3$. This large difference between the LO and the NLO
results for $\lambda_s$ is largely explained by the running of the
coupling, whose consequences are crucial for the physics of saturation.
The relevant value of the coupling is that corresponding to the
saturation momentum, $\alpha_s(Q_s(Y))\propto 1/\ln [Q_s^2(Y)/\Lam^2]$,
which decreases with $Y$. Hence, for sufficiently high energy, all the
other NLO corrections (like the NLO effects in the BFKL kernel
\cite{NLBFKL,Salam99}) are suppressed by $\alpha_s(Q_s(Y))\ll 1$, so that
the correct theory for saturation (at least for asymptotically large $Y$)
is the LO theory extended to running coupling.

The main effect of the running of the coupling is to considerably slow
down the evolution, with some dramatic consequences: \texttt{(i)} rather
than growing exponentially with $Y$, the saturation momentum grows only
like $\exp{\sqrt{Y}}$ \cite{SCALING,MT02}; \texttt{(ii)} the window
$\rho_{\rm geom}- \rho_s$ for geometric scaling outside saturation grows
very slowly, as $Y^{1/6}$ \cite{DT02}; \texttt{(iii)} for a nuclear
target, the dependence of $Q_s$ upon the atomic number $A$ is strongly
reduced by the running of coupling
--- for sufficiently high energy, there should be no difference between
the saturation scale of a nucleus and that of the proton \cite{AM03}.
This last feature has intriguing consequences for particle production in
proton--nucleus collisions at very high energies, perhaps at LHC
\cite{IIT04}.

Another important, and rather surprising, consequence of the running of
the coupling \cite{Dumitru:2007ew} is to improve the applicability of the
mean field approximation, that is, the BK equation. We have already
mentioned, at the end of Sect. 4, that the basic equations of the CGC
theory (Balitsky--JIMWLK and BK) ignore the correlations induced via
gluon number fluctuations. These correlations refer to the fact that the
gluons produced by the evolution can have common ancestors. Or it turns
out that, for a {\em fixed} coupling, these correlations do significantly
affect the picture of saturation \cite{IM032,MS04,IMM04,IT04} :  Although
they are mostly produced in the dilute regime, i.e., in the tail of the
gluon distribution at high $Q^2$, these correlations are rapidly
amplified by the BFKL evolution, so they eventually influence the
approach towards saturation. (In terms of diagrams, the evolution with
both fluctuations and saturation contains Pomeron loops.) Accordingly,
this evolution becomes stochastic and the saturation scale itself becomes
a {\em random variable}, whose dispersion increases with $Y$ : with a
fixed coupling, this rise is so fast that already for moderate values of
$Y$ it completely washes out the mean field picture (and, in particular,
the property of geometric scaling). Direct calculations of such effects
in QCD are extremely difficult (the complete theory for QCD evolution
with Pomeron loops is still lacking; see, however, Refs.
\cite{IT04,MSW05,KL05,BREM,Balit05,HIMST06}), but the effects of the
fluctuations can be appreciated from the experience with the
reaction--diffusion problem in statistical physics, and also with some
QCD--inspired models which allow for explicit numerical calculations and
belong to the universality class of reaction--diffusion
\cite{LL05,GS05,EGBM05,Iancu:2006jw,Munier:2008cg}. But these studies
also allow for the inclusion of a running coupling, and the effects of
that turn out to be dramatic \cite{Dumitru:2007ew} : the fluctuations are
strongly suppressed up to the highest values of $Y$ of interest, so that
the complete, stochastic, evolution gives essentially the same results as
its mean field approximation. This is to be attributed to the fact that,
with running coupling, the saturation front has a different shape (due to
the shrinking of the window for geometric scaling), which disfavors
fluctuations \cite{Dumitru:2007ew}. We thus conclude that the actual
evolution in QCD at high energy is not in the universality class of
reaction--diffusion, and this is somehow fortunate as it allows us to
rely on mean field approximations like the BK equation (with running
coupling, of course).

\section{Saturation models}

\begin{figure*}[t]
\centerline{\includegraphics[width=0.93\textwidth]{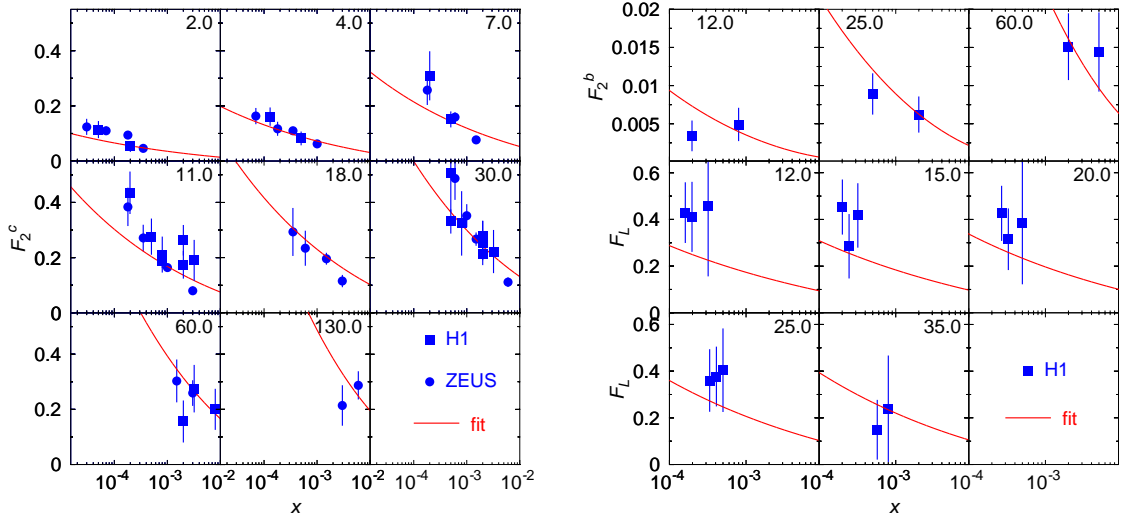}}
\vspace*{-.6cm} \caption{\sl Predictions of the `CGC model' with 4 quarks
(3 light + 1 heavy). Left: the charm structure function. Right: the
bottom structure function and the longitudinal structure function. From
Ref. \cite{GS07}.}\label{fig:othersf}
\end{figure*}

Although fully consistent with the gross features of the experimental
results at both HERA and RHIC, the current formalism for high--energy
evolution with saturation is not accurate enough to allow for a precise,
parameter--free, description of the small--$x$ HERA data, which are known
with high precision. The theoretical limitations refer to both
perturbative and non--perturbative aspects: on the one hand, the NLO
corrections (expected to be large) have not been systematically
implemented; on the other hand, the impact--parameter dependence of the
scattering amplitudes goes beyond perturbation theory (especially in the
dilute region towards the periphery of the hadron disk). To cope with
that, various ``saturation models'' have been proposed for the dipole
cross--section, Eq.~(\ref{sigdip}), which were inspired by theoretical
ideas about saturation, or by results from the BK equation, but which are
also involving several (typically, 3 or 4) free parameters. Such models
provided remarkably good descriptions of the relevant data at HERA and
RHIC --- even surprisingly good, given the simplicity of the models and
the reduced number of free parameters. Some general remarks about these
models: \texttt{(i)} The free parameters are fixed from fits to the $F_2$
data alone; all the other results emerge as {\em predictions}, and they
provide a reasonably good description of the {\em ensemble} of the HERA
data at $x\le 0.01$, including the longitudinal ($F_L$), diffractive
($F_2^D$), and charm ($F_2^c$) structure functions, the virtual photon
production of vector mesons ($\rho,\,J/\psi$), and the deeply virtual
Compton scattering (DVCS). \texttt{(ii)} The saturation models provide
natural explanations for important, qualitative, features of the data
like geometric scaling, the turn-over in $F_2$ at low $x$ and low $Q^2$,
and the nearly constant diffractive-to-inclusive ratio $\sigma_{\rm
diff}/\sigma_{\rm tot}$ at HERA, or the total multiplicity and the
high--$p_\perp$ suppression of particle production in forward d--Au
collisions (`$R_{pA}$ ratio') at RHIC. So far, there are no other
compelling explanations which apply to the {\em ensemble} of these data.

\begin{figure*}[t]
\centerline{
\includegraphics[width=0.5\textwidth]{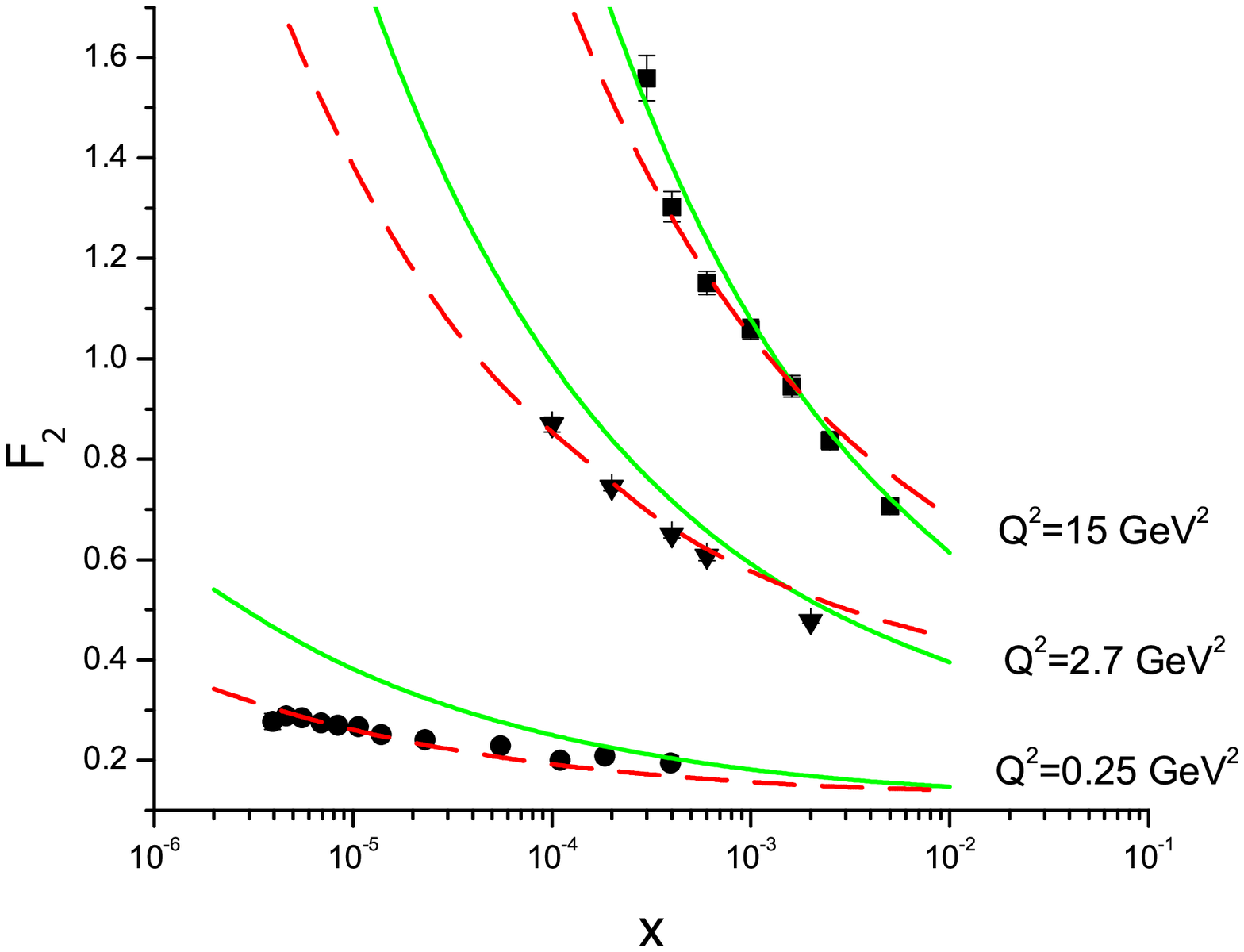}
\includegraphics[width=0.5\textwidth]{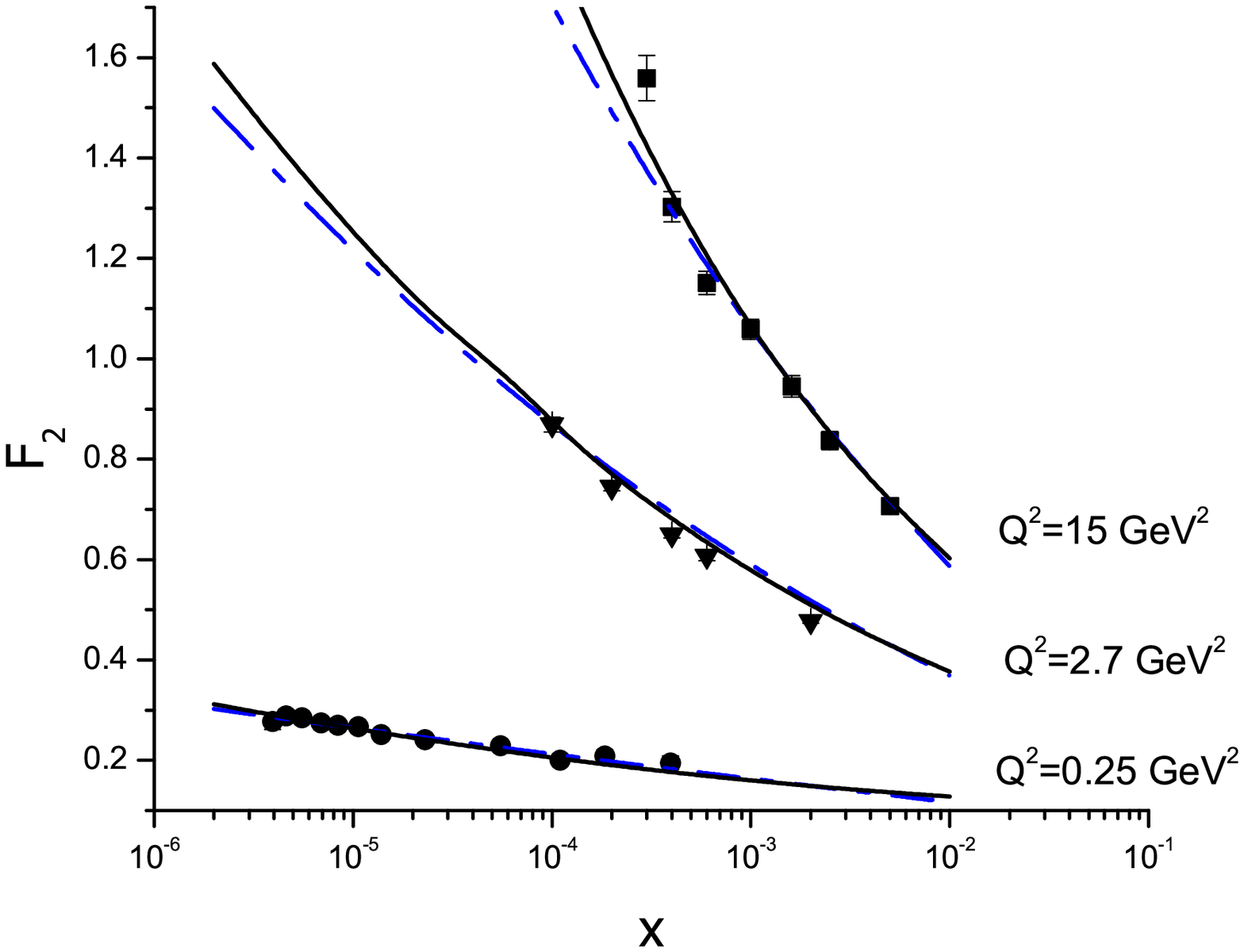}}\vspace*{-1.cm}
\caption{\sl Comparison of the various fits in Ref.~\cite{Forshaw2} to a
subset of DIS data. Left: No saturation fits. FS2004 Regge dipole fit
(dashed line) and (solid line) a fit of the same model to data in the
restricted range $5 \times 10^{-4} < x < 10^{-2}$, extrapolated over the
whole $x$-range $x < 0.01$. Right: Saturation fits. FS2004 saturation fit
(solid line) and the CGC dipole model (dot-dashed line)} \label{f2fits}
\end{figure*}

The first ``saturation model'', due to Golec-Biernat and W\"usthoff (GBW)
\cite{GBW99}, played an important role towards the shift of paradigm in
favour of saturation at HERA. The main virtue of that fit was in its
simplicity: with a very simple functional form,
 \beq
\sigma^{\rm GBW}_{\rm
 dipole}(x,r)&=&2\pi R^2\Big(
 1\,-\,{\rm e}^{-r^2 Q_s^2(x)}\Big),\nn Q_s^2(x) &=&(x_0/x)^\lambda\,
 {\rm GeV}^2\,,
 \eeq
which interpolates between color transparency ($\sigma_{\rm
dipole}\propto r^2$) for small dipole sizes and saturation ($\sigma_{\rm
dipole}\simeq 2\pi R^2$) for larger dipoles, and the transition occurring
at a `critical' scale $r_s\sim 1/Q_s(x)$ which decreases with $1/x$ (the
real hallmark of saturation), this model offered a good description of
the early HERA data (at $x\le 0.01$ and any $Q^2$) with only 3 free
parameters: the proton radius $R$, the value $x_0$ where $Q_s=1$ GeV, and
the saturation exponent $\lambda$ (the data favoured $x_0\approx 10^{-4}$
and $\lambda\approx 0.3$). Note that this model has exact geometric
scaling built in, and in fact it was his success which inspired the
search for this scaling in the data \cite{geometric}.

However, the limitations of this model become obvious with the advent of
more precise HERA data, and new, more sophisticated, models were then
proposed to account for these data. The main improvements referred to a
better inclusion of the effects of the perturbative QCD evolution (which
in particular brought in {\em violations} of geometric scaling),
sometimes accompanied by a more complex treatment of the
parameter--impact dependence. Some approaches \cite{BGBK,KT,KMW} focused
on improving the high--$Q^2$ behaviour of the fit (i.e., the small--$r$
behaviour of the dipole cross--section), by adding in the DGLAP
evolution. Some others have rather emphasized the BFKL physics and the
transition from linear to non--linear dynamics \cite{IIM03,GS07}.

In particular, the `CGC model' in Ref. \cite{IIM03}, which is based on
approximate solutions to the BK equation, has shown that both the BFKL
value for the `anomalous dimension' (the slope $\gamma \approx 0.63$ in
Eq.~(\ref{scaling})) and the pattern of geometric scaling violations
predicted by the BFKL diffusion are consistent with the HERA data. This
fit, which involves 3 light quarks and the same 3 free parameters as the
GBW fit, has also shown that the data prefer a smaller value for the
saturation exponent, namely $\lambda\approx 0.25$. This value has been
further reduced, to $\lambda\approx 0.22$, after also including the heavy
charm quark in the fit \cite{GS07}; this last analysis requires a
somewhat larger value $\gamma \approx 0.76$ for the `anomalous
dimension'. Some predictions of \cite{GS07} are summarized, together with
the respective data at HERA, in Fig.~\ref{fig:othersf}. The CGC model has
been recently extended to include impact--parameter dependence
\cite{KMW,Watt:2007nr}, but the respective fit favors an unusually small
value $\gamma \approx 0.46$ for the `anomalous dimension', which looks
inconsistent with the BK dynamics.

\begin{figure*}[t]
    \centerline{\includegraphics[width=0.9\textwidth]{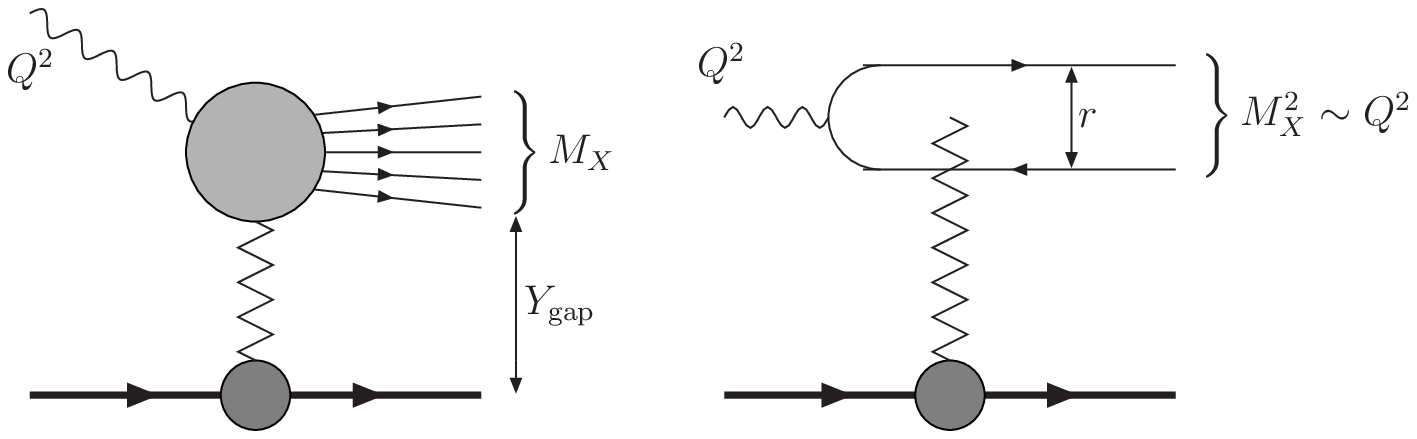}
}\vspace*{-.6cm}\caption{\sl Left: Schematic view of DIS diffraction.
Right : Quasi--elastic diffraction ($Y_{\rm gap}\simeq
Y$).}\label{fig:diff}
    \end{figure*}

An alternative saturation model, `FS2004', has been proposed in Refs.
\cite{Forshaw,Forshaw2,Forshaw3}, which is particularly simple (and thus
closer in spirit to the original GBW model), but also more flexible, in
the sense of including more free parameters. Interestingly, this fit has
two versions (with and without saturation), and the version including
saturation appears to be clearly favored \cite{Forshaw2} by the HERA data
for $F_2$, so long as all the data (including those at low $Q^2$) are
included in the fit. This is illustrated in Fig.~\ref{f2fits}
\cite{Forshaw2}, which also show that the quality of the FS2004 fit with
saturation is similar to that of the CGC fit in Ref.~\cite{IIM03}.

One should also mention here the saturation models used in the context of
RHIC, where the dipole cross--section (or, more precisely, its Fourier
transform) enters the rate for forward particle production in
deuteron--gold collisions. In that case, it is preferable to formulate
the models directly in momentum space, to avoid numerical artifacts
associated with the Fourier transform. The saturation models formulated
in Refs.~\cite{Kharzeev:2004yx,Dumitru:2005gt} provide a reasonable
description of all the relevant RHIC data; moreover, as shown in
Ref.~\cite{Goncalves:2006yt}, a slight modification of these models can
also account for the HERA data, as expected on the basis of the
universality of saturation physics.

\begin{figure}[t]
\begin{center}
    \centerline{\includegraphics[width=0.5\textwidth]{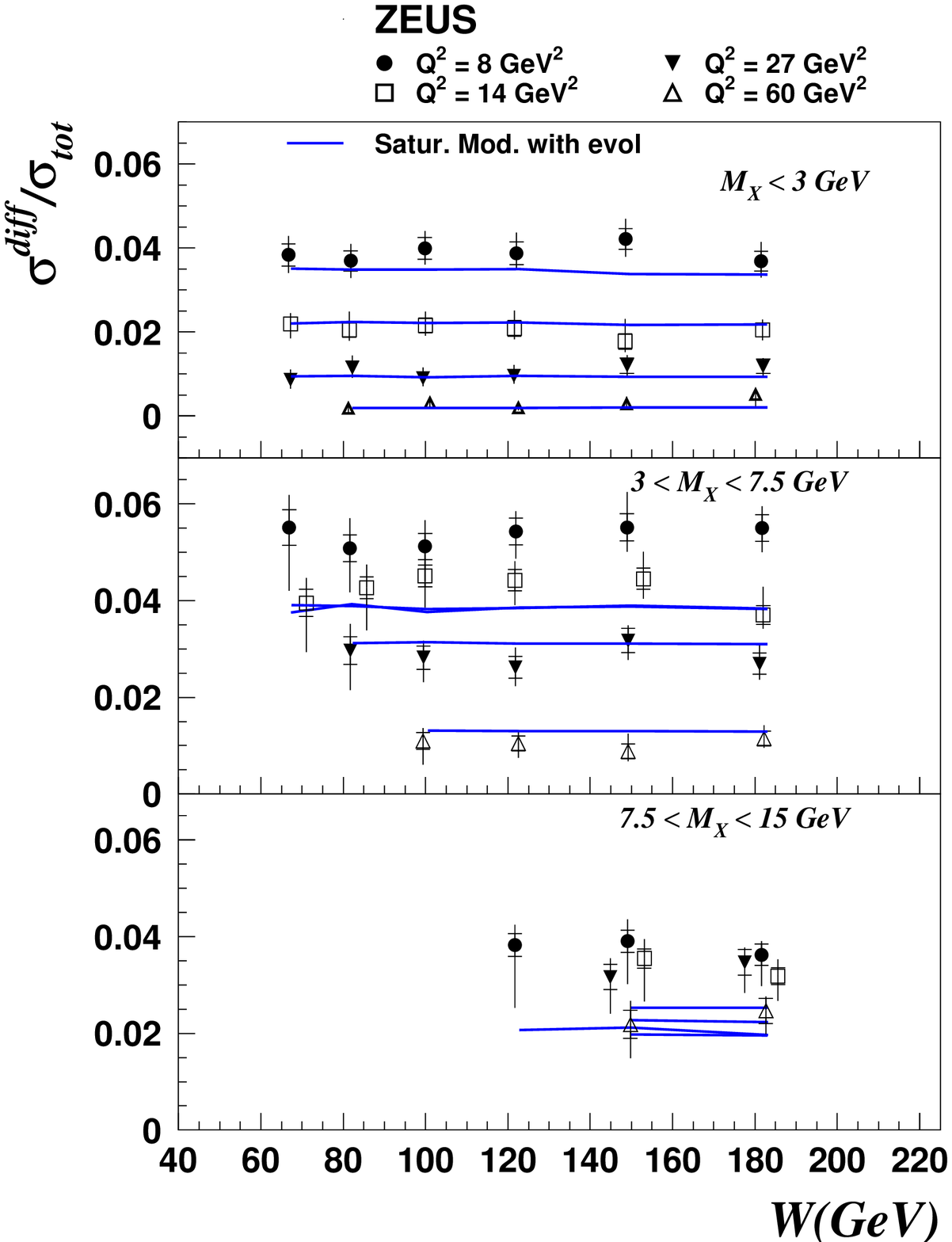}
} \vspace*{-.6cm}\caption{\sl ZEUS data for the ratio
$\sigma_{\rm diff}/\sigma_{\rm tot}$ together with the respective
prediction of the saturation model in Ref.~\cite{BGBK}.
}\label{fig:ratio}
 \end{center}\vspace*{-1.cm}
    \end{figure}

Returning to HERA physics, we shall conclude this review with a brief
discussion of {\em DIS diffraction}, which is a particularly convenient
laboratory to test saturation. Indeed, the diffractive cross--section is
controlled by large dipole sizes and hence it is particularly sensitive
to our theoretical ideas about unitarization. The theory is simpler for
the case of a large rapidity gap $Y_{\rm gap}\simeq Y$, or small
diffractive mass $M_X^2\sim Q^2$, in which case diffraction amounts to
the elastic scattering of the $q\bar q$ dipole produced by the
dissociation of $\gamma^*$ (see Fig.~\ref{fig:diff}). The respective
cross--section is then evaluated as (compare to Eq.~(\ref{dipolefact}))
 \beq
 \frac{\rmd\sigma_{\rm diff}}
 {\rmd^2 {b}}\,=\,
 \int\rmd z \, \rmd^2 {\bm r}\
 \vert \Psi_{\gamma}(z, r; Q)\vert^2 \ \, 
 \big( T(r, Y) \big)^2\,.\eeq
The photon wavefunction favors relatively small dipoles with $r\sim 1/Q$
:
 \beq\label{intdiff}
 \frac{\rmd\sigma_{\rm diff}}
 {\rmd^2 {b}}
 \,\sim\,
 \,\frac{1}{Q^2}\int\limits_{1/Q^2}^{\infty}
 \frac{\rmd r^2 }{r^4}\,\big( T(r, Y) \big)^2\,.\eeq
But for small $r$, $T(r,Y) \propto r^2$, hence the integral will be
dominated by the size $r_s$ where the amplitude has a turn--over, due to
unitarity corrections. If unitarization is to be associated with the
soft, non--perturbative, physics (the prevailing viewpoint before the
advent of saturation; see, e.g., \cite{Bjorken:1996vu}), then $r_s\sim
1/\Lam$, and diffraction would be non--perturbative even when $Q^2$ is
hard~! However, for sufficiently large $Y$ (small $x$), the (semi)hard
saturation scale $Q_s(Y)$ enters the game and cuts off the integral in
Eq.~(\ref{intdiff}):
 \beq\label{diffsat}
 \frac{\rmd\sigma_{\rm diff}}
 {\rmd^2 {b}}
\sim
 \,\frac{1}{Q^2}\int\limits_{1/Q^2}^{1/Q_s^2}
 \frac{\rmd r^2 }{r^4}\, \Big(r^2Q_s^2(x)\Big)^2
  \, \sim \, \frac{Q_s^2(x)}{Q^2}\,. 
 \eeq
In this scenario, $\sigma_{\rm diff}\propto Q_s^2(x) \propto
x^{-\lambda}$ scales like the `hard Pomeron', that is, in the same way as
the inclusive cross--section (\ref{dipolefact}); hence the ratio
$\sigma_{\rm diff}/\sigma_{\rm tot}$ is approximately constant as a
function of the energy. This prediction is to be contrasted to that of
non--perturbative unitarization \cite{Bjorken:1996vu}, where one rather
expects the diffractive cross--section to rise twice as fast as the gluon
distribution (since elastic scattering requires the exchange of at least
two gluons), which would imply $\sigma_{\rm diff}/\sigma_{\rm tot}\sim
x^{-\lambda}$. It turns out that the HERA data favor the saturation
scenario: the measured ratio $\sigma_{\rm diff}/\sigma_{\rm tot}$ is very
flat as a function of the invariant energy $W^2\propto 1/x$, as
illustrated in Fig.~\ref{fig:ratio}. This figure also shows that this
flatness is well reproduced by the saturation model of Ref.~\cite{BGBK}.
Notice that the integrand in Eq.~(\ref{diffsat}) rises as a double (hard)
Pomeron, $Q_s^4(x) \propto x^{-2\lambda}$, but one power of $Q_s^2(x)$ is
eventually compensated by the energy--dependence of the upper cutoff
$r_s^2=1/Q_s^2(x)$, which is the critical size for the onset of unitarity
in the framework of saturation.

\subsection*{Acknowledgments}
I would like to thank the organizers of the Ringberg Workshop ``New
Trends in HERA Physics 2008'' for their warm hospitality at the Ringberg
Castle. This work is supported in part by Agence Nationale de la
Recherche via the programme ANR-06-BLAN-0285-01.


\begin{thebibliography}{10}

\bibitem{BFKL}
L.N.~Lipatov, {\it Sov.\ J.\ Nucl.\ Phys.}\,{\bf 23} (1976) 338;
E.A.~Kuraev,  L.N.~Lipatov and V.S.~Fadin, {\it Zh. Eksp. Teor. Fiz}
{\bf 72}, 3 (1977); Ya.Ya.~Balitsky,
  L.N.~Lipatov, {\it Sov.\ J.\ Nucl.\ Phys.} {\bf 28} (1978) 822.

\bibitem{GLR}
L.V.~Gribov, E.M.~Levin, and M.G.~Ryskin, {\it Phys. Rept. } {\bf
100} (1983)
  1.

\bibitem{MQ85}
A.H.~Mueller and J.~Qiu, {\it Nucl. Phys.} {\bf B268} (1986) 427.

\bibitem{DGLAP}
V.N.~Gribov and L.N.~Lipatov, {\it Sov.\ Journ.\ Nucl.\ Phys.}\ {\bf 15}
 (1972), 438; G. Altarelli and G. Parisi,
 {\it Nucl.\ Phys.}\,{\bf B126} (1977), 298;
  Yu. L.~Dokshitzer, {\it Sov.\ Phys.\ JETP} {\bf 46} (1977), 641.

\bibitem{MV}
L.~McLerran and R.~Venugopalan, {\it Phys.\ Rev.}\ {\bf D49} (1994)
2233; {\it
  ibid.} {\bf 49} (1994) 3352; {\it ibid.} {\bf 50} (1994) 2225.

\bibitem{JKLW} J.~Jalilian-Marian, A.~Kovner, A.~Leonidov and H.~Weigert,
    {\it Nucl.\ Phys.}\
  {\bf B504} (1997) 415; {\it Phys.\ Rev.}\ {\bf D59} (1999) 014014;
  J.~Jalilian-Marian, A.~Kovner and H.~Weigert, {\it Phys.\ Rev.}\ {\bf D59}
  (1999) 014015; A. Kovner, J. G. Milhano and H. Weigert, {\it Phys. Rev.} {\bf
  D62} (2000) 114005; H.~Weigert, {\it Nucl. Phys.} {\bf A703} (2002) 823.

\bibitem{CGC} E.~Iancu, A.~Leonidov and L.~McLerran, {\it Nucl.
    Phys.}~{\bf A692} (2001) 583;
  {\it Phys. Lett.} {\bf B510} (2001) 133; E.~Ferreiro, E.~Iancu, A.~Leonidov
  and L.~McLerran, {\it Nucl. Phys.} {\bf A703} (2002) 489.

\bibitem{B} I.~Balitsky, {\it Nucl.\ Phys.}\ {\bf B463} (1996) 99; {\it
    Phys. Lett.} {\bf
  B518} (2001) 235.

\bibitem{K} Yu.V.~Kovchegov, {\it Phys. Rev.} {\bf D60} (1999) 034008;
    {\it ibid.} {\bf
  D61} (1999) 074018.


\bibitem{EdiCGC}
E.~Iancu, A.~Leonidov and L.~McLerran, hep-ph/0202270;
 E.~Iancu and R.~Venugopalan, hep-ph/0303204; H. Weigert,
  hep-ph/0501087.

\bibitem{geometric} A. Stasto, K. Golec-Biernat and J.~Kwiecinski, {\it
    Phys. Rev. Lett.} {\bf 86} (2001) 596.

\bibitem{MS06} C. Marquet and L. Schoeffel, {\em Phys.\ Lett.\ } {\bf
    B639}
    (2006) 471,
    hep-ph/0606079.

\bibitem{Brahms-data}
I.~Arsene {\it et al.} [BRAHMS Collaboration], {\it Phys.\ Rev.\
Lett.}\ {\bf
  93} (2004) 242303.



\bibitem{NLBFKL}
V.S.~Fadin and L.N.~Lipatov, {\it Phys. Lett.} {\bf B429} (1998)
127; G. Camici
  and M. Ciafaloni, {\it Phys. Lett.} {\bf B430} (1998) 349.

\bibitem{Salam99} G.P. Salam, {\em JHEP} {\bf 9807} (1998) 19; M.
    Ciafaloni, D.
    Colferai, {\it Phys.
  Lett.} {\bf B452} (1999) 372; M. Ciafaloni, D. Colferai, and G.P. Salam, {\it
  Phys. Rev. } {\bf D60} (1999) 114036.

\bibitem{SAT}
E.~Iancu and L.~McLerran, {\it Phys. Lett.} {\bf B510} (2001) 145.

\bibitem{AM02}
A. H. Mueller, {\it Nucl. Phys.} {\bf B643} (2002) 501.

\bibitem{GAUSS}
E.~Iancu, K.~Itakura, and L.~McLerran, {\it Nucl. Phys.} {\bf A724}
(2003) 181.

\bibitem{AM99}
A. H. Mueller, {\it Nucl. Phys.} {\bf B558} (1999) 285.



  \bibitem{Kovchegov:2008mk}
  Y.~V.~Kovchegov, J.~Kuokkanen, K.~Rummukainen and H.~Weigert,
  arXiv:0812.3238

\bibitem{JalilianMarian:2005jf}
  J.~Jalilian-Marian and Y.~V.~Kovchegov,
  {\em Prog.\ Part.\ Nucl.\ Phys.\ }  {\bf 56} (2006) 104.

\bibitem{Gelis:2007kn}
  F.~Gelis, T.~Lappi and R.~Venugopalan,
 {\em Int.\ J.\ Mod.\ Phys.\ }  {\bf E16} (2007) 2595.

  \bibitem{CGCRHIC}
M. Gyulassy, L. McLerran,  {\it Nucl.~Phys.~}{\bf A750} (2005) 30; J.-P.
Blaizot, F. Gelis, {\it ibid.} 148.


 \bibitem{Abreu:2007kv}
  N.~Armesto {\it et al.},
  {\em J.\ Phys.\  } {\bf G35} (2008) 054001
  [arXiv:0711.0974 [hep-ph]].

\bibitem{KM98} Yu. V. Kovchegov and A. H. Mueller, {\it Nucl. Phys.} {\bf
    B529} (1998) 451; Yu. V. Kovchegov and K. Tuchin,
    {\it Phys. Rev.} {\bf D65} (2002) 074026.

\bibitem{CM04} C. Marquet, {\it Nucl. Phys.} {\bf B705} (2005) 319.

\bibitem{Gelis:2008rw}
  F.~Gelis, T.~Lappi and R.~Venugopalan,
  arXiv:0804.2630; arXiv:0807.1306 [hep-ph].

  \bibitem{Balitsky:2006wa}
  I.~Balitsky,
  {\em Phys.\ Rev.\  } {\bf D75} (2007) 014001;
  I.~Balitsky and G.~A.~Chirilli,
  {\em Phys.\ Rev.\ } {\bf D77} (2008) 014019.


\bibitem{Kovchegov:2006vj}
  Y.~V.~Kovchegov and H.~Weigert,
  {\em Nucl.\ Phys.\ }  {\bf A784} (2007) 188;
   J.~L.~Albacete and Y.~V.~Kovchegov,
  {\em Phys.\ Rev.\  } {\bf D75} (2007) 125021.

  \bibitem{AM03} A.H.~Mueller, Nucl.\ Phys.\  A {\bf 724} (2003)
    223.

\bibitem{IIT04} E.~Iancu, K.~Itakura, and D.N.~Triantafyllopoulos, {\it
    Nucl. Phys.} {\bf A742}
  (2004) 182.

\bibitem{IM032} E.~Iancu and A.H.~Mueller, {\it Nucl.\ Phys.}\ {\bf A730}
    (2004) 494.

\bibitem{MS04} A.H.~Mueller and A.I.~Shoshi, {\it Nucl.\ Phys.}\ {\bf
    B692} (2004) 175.

\bibitem{IMM04} E.~Iancu, A.H.~Mueller and S.~Munier, {\it
    Phys.~Lett.~}{\bf B606} (2005) 342.

\bibitem{IT04} E.~Iancu and D.N.~Triantafyllopoulos, {\it
    Nucl.~Phys.~}{\bf A756} (2005) 419; {\it Phys.~Lett.~}{\bf B610}
(2005) 253.


\bibitem{Dumitru:2007ew}
  A.~Dumitru, E.~Iancu, L.~Portugal, G.~Soyez and D.~N.~Triantafyllopoulos,
 {\em JHEP} {\bf 0708} (2007) 062.

\bibitem{SCALING} E.~Iancu, K.~Itakura, and L.~McLerran, {\it Nucl.
    Phys.} {\bf A708} (2002) 327.

\bibitem{MT02} A.H.~Mueller and D.N.~Triantafyllopoulos, {\it Nucl.
    Phys.} {\bf B640} (2002)  331.

\bibitem{MP03} S.~Munier and R.~Peschanski, {\it Phys. Rev. Lett.} {\bf
    91} (2003) 232001.

\bibitem{Saar} For a review, see W.~Van Saarloos, {\it Phys. Rep.} {\bf
    386} (2003) 29.

\bibitem{DT02}
D.N.~Triantafyllopoulos, {\it Nucl. Phys.} {\bf B648} (2003) 293.

\bibitem{IIM03}
E.~Iancu, K.~Itakura and S.~Munier, {\it Phys. Lett.} {\bf B590}
(2004) 199.

\bibitem{GS07}
  G.~Soyez,
 {\em Phys.\ Lett.\ } {\bf B655} (2007) 32.


\bibitem{MSW05} A.H.~Mueller, A.I.~Shoshi, S.M.H.~Wong, {\it
    Nucl.~Phys.~}{\bf B715} (2005)
  440.

\bibitem{KL05} A.~Kovner, M.~Lublinsky, {\it Phys.~Rev.~}{\bf D71} (2005)
    085004; {\it
  Phys.~Rev.~Lett.~}{\bf 94} (2005) 181603.

\bibitem{BREM} Y. Hatta, E. Iancu, L. McLerran, A. Stasto, and
    D.N.~Triantafyllopoulos, {\it
  Nucl. Phys.} {\bf A764} (2006) 423.


\bibitem{Balit05} I. Balitsky, {\it Phys. Rev.} {\bf D72} (2005) 074027.


\bibitem{HIMST06} Y.~Hatta, E.~Iancu, C. Marquet, G. Soyez, and
    D.N.~Triantafyllopoulos, {\it  Nucl. Phys.} {\bf A773} (2006) 95.

\bibitem{LL05} E.~Levin and M.~Lublinsky, {\it Nucl.~Phys.~}{\bf A763}
    (2005) 172.

\bibitem{GS05}
G. Soyez, {\it Phys. Rev.} {\bf D72} (2005) 016007.

\bibitem{EGBM05}
R. Enberg, K. Golec--Biernat, and S. Munier, {\it Phys. Rev.} {\bf
D72} (2005)
  074021.

\bibitem{Iancu:2006jw}
  E.~Iancu, J.~T.~de Santana Amaral, G.~Soyez and D.~N.~Triantafyllopoulos,
 {\em Nucl.\ Phys.\ }  {\bf A786} (2007) 131.

\bibitem{Munier:2008cg}
  S.~Munier, G.~P.~Salam and G.~Soyez,
  arXiv:0807.2870 [hep-ph].

\bibitem{GBW99} K.~Golec-Biernat and M.~W{\"{u}}sthoff, {\em Phys. Rev. }
    \textbf{D59} (1999) 014017; {\em Phys. Rev. }
    \textbf{D60} (1999) 114023.

    \bibitem{BGBK}
J. Bartels, K. Golec-Biernat, and H. Kowalski, {\it Phys. Rev.} {\bf D66}
  (2002) 014001.

\bibitem{KT} H.~Kowalski and D.~Teaney, {\it Phys. Rev.} {\bf D68} (2003)
    114005.

    \bibitem{KMW}
  H.~Kowalski, L.~Motyka and G.~Watt,
  {\em Phys.\ Rev.\ } {\bf D74} (2006) 074016.

  \bibitem{Watt:2007nr}
  G.~Watt and H.~Kowalski,
  {\em Phys.\ Rev.\ } D {\bf 78} (2008) 014016.

\bibitem{Forshaw} J.~R.~Forshaw, R.~Sandapen and G.~Shaw, {\it Phys.
    Rev.} {\bf D69} (2004)
  094013; {\it Phys. Lett.} {\bf B594} (2004) 283.

\bibitem{Forshaw2}
J.~R.~Forshaw and G.~Shaw, 
 {\em JHEP } {\bf 0412}, 052 (2004).

  \bibitem{Forshaw3}
  J.~R.~Forshaw, R.~Sandapen and G.~Shaw,
  {\em JHEP } {\bf 0611} (2006) 025.

  \bibitem{Kharzeev:2004yx}
  D.~Kharzeev, Y.~V.~Kovchegov and K.~Tuchin,
  {\em Phys.\ Lett.\ } {\bf B599} (2004) 23.

  \bibitem{Dumitru:2005gt}
  A.~Dumitru, A.~Hayashigaki and J.~Jalilian-Marian,
  {\em Nucl.\ Phys.\ } A {\bf A765} (2006) 464.

\bibitem{Goncalves:2006yt}
  V.~P.~Goncalves, M.~S.~Kugeratski, M.~V.~T.~Machado and F.~S.~Navarra,
  {\em Phys.\ Lett.\ } {\bf B643} (2006) 273.

  \bibitem{Bjorken:1996vu}
  J.~D.~Bjorken,
  arXiv:hep-ph/9601363.

\end{thebibliography}
\end{document}